\documentclass{article}
\usepackage{amsmath, amssymb}

\usepackage[square, numbers]{natbib}

\usepackage{hyperref, hypernat}

\usepackage{graphicx, subcaption}
\usepackage{authblk}
\usepackage[margin=1.1in]{geometry}

\graphicspath{ {images/} }

\begin{document}

\title{Spinsim: a GPU optimized python package for simulating spin-half and spin-one quantum systems}

\author[1]{Alex Tritt}
\author[1, 3]{Joshua Morris}
\author[1, 4]{Joel Hochstetter}
\author[5, 6]{R. P. Anderson}
\author[2]{James Saunderson}
\author[1]{L. D. Turner}

\affil[1]{School of Physics \& Astronomy, Monash University, Victoria 3800, Australia.}
\affil[2]{Department of Electrical and Computer Systems Engineering, Monash University, Victoria 3800, Australia.}
\affil[3]{Vienna Center for Quantum Science and Technology (VCQ), Faculty of Physics, University of Vienna, 1010 Vienna, Austria.}
\affil[4]{Department of Applied Mathematics and Theoretical Physics, Centre for Mathematical Sciences, University of Cambridge, Cambridge, UK.}
\affil[5]{School of Molecular Sciences, La Trobe University, PO box 199, Bendigo, Victoria 3552, Australia.}
\affil[6]{Q-CTRL Pty. Ltd}

\maketitle{}

\begin{abstract}
	The \emph{Spinsim} \emph{python} package simulates spin-half and spin-one quantum mechanical systems following a time dependent Shr\"odinger equation.
	It makes use of \texttt{numba.cuda}, which is an \emph{LLVM} (Low Level Virtual Machine) compiler for \emph{Nvidia Cuda} compatible systems using GPU parallelization. 
	Along with other optimizations, this allows for speed improvements from 3 to 4 orders of magnitude while staying just as accurate, compared to industry standard packages. 
	It is available for installation on \emph{PyPI}, and the source code is available on \emph{github}.
	The initial use-case for the \emph{Spinsim} will be to simulate quantum sensing-based ultracold atom experiments for the Monash University School of Physics \& Astronomy spinor Bose-Einstein condensate (spinor BEC) lab, but we anticipate it will be useful in simulating any range of spin-half or spin-one quantum systems with time dependent Hamiltonians that cannot be solved analytically.
	These appear in the fields of nuclear magnetic resonance (NMR), nuclear quadrupole resonance (NQR) and magnetic resonance imaging (MRI) experiments and quantum sensing, and with the spin-one systems of nitrogen vacancy centres (NVCs), ultracold atoms, and BECs.
\end{abstract}

\section{Overview}
	\subsection{Introduction}
		\subsubsection{Motivation}
			Ultracold rubidium atoms have proven their effectiveness in state of the art technologies in quantum sensing~\cite{degen_quantum_2017}, the use of quantum mechanics to make precise measurements of small signals.
			These atoms can be modelled as quantum spin systems, which is the quantum mechanical model for objects with angular momentum.
			The simplest spin system, spin-half (i.e. spin quantum number of $ \frac12 $, also referred to as a qubit), is quantized into just two quantum spin levels, and this describes the motion of some fundamental particles such as electrons.
			However, systems more practical for sensing, such as ultracold rubidium atoms, are more accurately described as a spin-one quantum system (i.e. spin quantum number of $ 1 $, also referred to as a qutrit), which is quantized into three quantum spin levels.
			
			The design of sensing protocols requires many steps of verification, including simulation.
			This is especially important, since running real experiments can be expensive and time consuming, so it is more practical to debug such protocols quickly and cheaply on a computer.
			In general, any design of experiments using spin systems could benefit from a fast, accurate method of simulation.

			In the past, the spinor Bose Einstein condensate (spinor BEC) lab at Monash University used an in-house, \emph{cython} based script, on which this package, \emph{Spinsim}, is based, and standard differential equation solvers (such as \emph{Mathematica}'s~\cite{wolfram_research_inc_mathematica_2020} function \texttt{NDSolve}) to solve the Schr\"odinger equation for quantum sensing spin systems.
			However, these methods are not completely optimized for our use-case, and therefore come with some issues.

			First, while the execution time for these solvers is acceptable for running a small number of experiments, for certain experiments involving large arrays of independent atom clouds (which require many thousands of simulations to be run), this time accumulates to the order of many hours, or even multiple days.
			A crude method often used to combat this involves approximating spin-one systems as spin-half for a faster execution time, at the cost of not modelling all effects in the system.
			With \emph{Spinsim}, this is something we would like to avoid.


			One option not explored by these generic methods is parallelization, specifically across graphics processing unit (GPU) cores.
			While there do already exists parallelized algorithms to solve general differential equations~\cite{lions_resolution_2001}, there are properties specific to the spin system that allow for much more fast and elegant parallelization in our use-case.
			We use these properties (see Section {\ref{sec:parallelization}}) to write a parallel, GPU based solver that runs faster than industry standard non-parallel algorithms by up to 4 orders of magnitude.

			Second, the Schr\"odinger equation has the geometric property of being norm preserving.
			In other words, the time-evolution operator for a system between two points in time must be unitary.
			As such, numerical solutions to the Schr\"odinger equation should also preserve norms.
			For many numerical methods like those in the Runge Kutta family, the approximations used are in general not norm preserving, so the evaluated quantum state may diverge towards an infinite norm, or converge to zero if run for many iterations.
			To avoid this, we use a geometric integration technique for our solver (see Section {\ref{sec:magnus}}).

			Third, our system (and similar spin systems) can be very oscillatory.
			Standard integration methods require very small time-steps in order to accurately depict these oscillations.
			For this reason we use a fourth order Magnus expansion~\cite{magnus_exponential_1954,blanes_magnus_2009} based solver for our geometric integrator to make larger time-steps accurate (see Section {\ref{sec:magnus}}).
			We also use a dynamically changing rotating frame to slow down the effective rotations in the system, and thus reduce the size of integration steps being done, making each step more accurate (see Section {\ref{sec:rotating_frame}}).


		\subsubsection{Comparison to recent specialized packages} 
			While we benchmark \emph{Spinsim} only against industry standard integrators, we also note that application specific quantum numerical simulator packages are regularly being released.
			However, none of these packages are designed specific to quantum sensing, and so these releases are not suitable for simulations highly accurate at high time resolutions required in this field.
			See Section {\ref{sec:applications}} for use cases where \emph{Spinsim} is advantageous.

			One application of simulations of spin systems is nuclear magnetic resonance (NMR).
			In this field, the simulator part of the \emph{PULSEE}~\cite{candoli_pulsee_2021} package uses up to the 3rd order Magnus expansion just once to solve the complete system, rather than using it many times as a time stepping method.
			Nested integrals of commutators are numerically evaluated on a CPU using Euler integration.
			This is likely to be slow (due to the Euler-integrated nested integrals of commutators) and inaccurate (due to using the expansion over a large time step), and potentially not even convergent in general (also due to the large time step) for solving complicated time dependent systems in quantum sensing.
			With that said, this may not matter for the kinds of problems this NMR simulation package is looking at.
			This is an application that would benefit from \emph{Spinsim}'s speed.

			An application of spin system simulation that would benefit to a lesser extent is quantum optimal control (QOC).
			In QOC, researchers look at how to design robust pulse sequences to control spin systems using machine learning techniques.
			Particularly, this means that simulators used here require automatic differentiation~\cite{griewank_who_2012}, which \emph{Spinsim} does not currently support.
			A main use of QOC is to design control sequences in quantum computers.
			Many of these pulse sequences are trains of hard pulses (which are physically constant bursts of radiation), which, if using the accuracy reducing rotating wave approximation (RWA), can have simplified time evolution operators.
			This paradigm, used in the recent \emph{Qulacs}~\cite{suzuki_qulacs_2021}, \emph{QOpt}~\cite{teske_qopt_2021} and \emph{QSim}~\cite{isakov_simulations_2021}, is not suitable for accurately simulating more complicated (especially when looking from the lab frame of reference) pulse shapes found in quantum sensing.
			However, all of these packages do allow for simulating entanglement (important for quantum computing, but not for all sensing applications), which \emph{Spinsim} does not and, like \emph{Spinsim}, all have versions optimized for parallelization (although \emph{QSim} does sacrifice accuracy here for the sake of speed, by using single-precision floats).
			Dalgaard and Motzoi~\cite{dalgaard_fast_2021} have recently proposed Magnus integration techniques for use in QOC, but have not provided any open source implementation to do so.




	\subsection{Implementation and architecture}
		\subsubsection{Quantum mechanics background}
			The \emph{Spinsim} package solves the time-dependent Schr\"{o}dinger equation
			\begin{align}
				i\hbar\frac{\text{d}\psi(t)}{\text{d}t} &= H(t)\psi(t),\label{eq:schroedinger}
			\end{align}
			where the quantum state $ \psi(t) \in \mathbb{C}^N $ is assumed normalized, and the Hamiltonian $ H(t) \in \mathbb{C}^{N \times N} $ is Hermitian.
			Here $ N $ is the number of levels in the quantum system.
			Often we are considering systems with spin of half ($ N = 2 $ case) or one ($ N = 3 $).
			We set $ \hbar = 1 $, so that the Hamiltonian has physical dimension of angular frequency.
			
			

			Rather than parameterizing the problem in terms of the matrix elements of $ H(t) $, we consider time varying real coefficients in a linear combination of a basis of fixed operators,
			\begin{align}
				H(t) &= \sum_{j = 1}^{N^2 - 1} \omega_j(t) \mathcal{A}_j.
			\end{align}
			Here we exclude the identity so that the Hamiltonian is traceless, which corresponds to choosing a physically meaningless energy zero point.
			It is well known~\cite[(p74)]{j_j_sakurai_jun_john_modern_2011} that a charged spin-half system with magnetic moment $ \overrightarrow{\mu} $, and gyromagnetic ratio $ \gamma $, in a magnetic field $ \overrightarrow{B}(t) $, has Hamiltonian
			\begin{align}
				H(t) &= -\overrightarrow{B}(t)\cdot \overrightarrow{\mu}\\
				&= -\gamma \left(B_x(t) J_x + B_y(t) J_y + B_z(t) J_z\right)\\
				&= \omega_x(t) J_x + \omega_y(t) J_y + \omega_z(t) J_z.
			\end{align}
			Here $ J_x $, $ J_y $ and $ J_z $ are spin operators, equal to the Pauli matrices~\cite[(p169)]{j_j_sakurai_jun_john_modern_2011} halved.
			Motivated by this, we define our basis for the spin-half model with $ \mathcal{A}_1 = J_x $, $ \mathcal{A}_2 = J_y $ and $ \mathcal{A}_3 = J_z $.
			Additionally, because of the equivalence of the magnetic field components $ B_j(t) $ and the $ \omega_j(t) $, we henceforth refer to the latter as field functions.

			In the spin-one case there is no universal standard basis of operators $ \mathcal{A}_j $.
			Choices include the Gell-Mann matrices~\cite{gell-mann_symmetries_1962}, and multiple dipole-quadrupole bases~\cite{hamley_spin-nematic_2012, di_dipolequadrupole_2010}.
			In general, we can choose any basis from the 8-dimensional Lie algebra $ \mathfrak{su}(3) $, which is the vector space of traceless Hermitian operators that can generate transformations (from the corresponding Lie group $ SU(3) $) in the spin-one system.
			We focus on a particular subfamily of spin-one systems for which the Hamiltonian is a linear combination of matrices from a 4-dimensional subspace of $ \mathfrak{su}(3) $, consisting of the spin matrices $ J_x $, $ J_y $ and $ J_z $, (labelled $ L_x $, $ L_y $ and $ L_z $ in Reference~\cite{hamley_spin-nematic_2012}) and a single quadrupole operator $ Q = \text{diag}(1, -2, 1)/3 $,


			\begin{align}
				H(t) &= \omega_x(t) J_x + \omega_y(t) J_y + \omega_z(t) J_z + \omega_q(t) Q.
			\end{align}
			
			Note that $ Q $ is proportional to $ J_z^2 $ (up to a change in energy zero point), and $ Q_{zz} $~\cite{hamley_spin-nematic_2012} and $ Q_0 $~\cite{di_dipolequadrupole_2010} from alternative quadrupole bases.
			The $ J_x $, $ J_y $, $ J_z $ and $ Q $ are the only operators necessary to simulate many spin-one quantum systems in arbitrary bias fields, but with single-photon coupling.
			This includes quadratic Zeeman splitting described by the Breit-Rabi formula~\cite{mockler_atomic_1961}, important to experiments in our lab.
			The \emph{Spinsim} simulator can also be configured to solve a general spin-one system by setting the Hamiltonian to an arbitrary point in $ \mathfrak{su}(3) $ using the full quadrupole basis, which extends the possible Hamiltonian to
			\begin{align}
				H(t) =& \omega_x(t) J_x + \omega_y(t) J_y + \omega_z(t) J_z + \omega_q(t) Q\nonumber\\
				&+ \omega_{u1}(t) U_1 + \omega_{u2}(t) U_2 + \omega_{v1}(t) V_1 + \omega_{v2}(t) V_2.
			\end{align}
			Here the additional operators are those defined in~\cite{di_dipolequadrupole_2010}. Note that this is included for completeness, but has not been thoroughly tested, as our lab has no physical context for modelling the $ U_1, U_2, V_1 $ and $ V_2 $ operators.
			As well as integrating the Schr\"odinger equation, \emph{Spinsim} also has the functionality to calculate the expected spin projection $ \langle \overrightarrow{J}\rangle (t) $ of a system from its state.
			In an experimental setting, one cannot measure the value of the state itself, and must instead measure observables such as spin projection.
			If this is done for an ensemble of systems, the average spin projection over all systems will converge to the expected value as calculated here.
			As an alternative to solving the Schr\"odinger equation, if one is only interested in the dynamics of the expected spin projection, one can instead solve for $ \langle \overrightarrow{J}\rangle (t) $ directly using the Bloch equations~\cite{bloch_nuclear_1946}.
			These are three coupled real equations as compared to three coupled complex equations for the spin-one Schr\"odinger equation.
			However, the Bloch equations cannot be used to compute $ \psi(t) $, and can only fully model spin-half systems (and spin-one systems approximated to spin-half).
			Thus, with \emph{Spinsim} we choose to solve the full Schr\"odinger equation instead.
			


		
		\subsubsection{Unitary time evolution and the Magnus expansion}\label{sec:magnus}
			Since $H(t)$ is Hermitian, it follows from Equation~\eqref{eq:schroedinger} that $\|\psi(t)\|^2 = \|\psi(t_0)\|^2$ for all $t$.
			Therefore it is possible to write $ \psi(t) $ in terms of a unitary transformation $ \mathcal{U}(t, t_0) $ of the state $ \psi(t_0) $, for any time $ t_0 $, that is, $ \psi(t) = \mathcal{U}(t, t_0)\psi(t_0) $.
			It then follows from Equation~\eqref{eq:schroedinger} that $ \mathcal{U}(t, t_0) $ also follows a Schr\"{o}dinger equation,
			\begin{align}
				i\hbar\frac{\text{d}\mathcal{U}(t, t_0)}{\text{d}t} &= H(t)\mathcal{U}(t, t_0).\label{eq:unitary_schroedinger}
			\end{align}

			Thus, to solve Equation~\eqref{eq:schroedinger} for $ \psi(t) $ given a particular $ \psi(t_0) $, one only needs to solve Equation~\eqref{eq:unitary_schroedinger} for $ \mathcal{U}(t, t_0) $.
			Since $ \mathcal{U}(t, t_0) $ is unitary, it can be written as a matrix exponential of a anti-Hermitian (also termed skew-Hermitian) matrix.
			If the Hamiltonian is a of constant value $ H(t) = \mathcal{H} $, then this exponential is 
			\begin{align}
				\mathcal{U}(t, t_0) &= \exp(-i (t - t_0) \mathcal{H}).\label{eq:exp_sol_of_constant}
			\end{align}
			For a time varying Hamiltonian, the general solution for $ \mathcal{U}(t, t_0) $ is much more complex, because it encapsulates the full solution to the time-dependent Schr\"{o}dinger equation.
			The well known Dyson series~\cite{kalev_integral-free_2020} gives an explicit expression for $ \mathcal{U}(t, t_0) $ in terms of multiple time integrals over nested time commutators of $ H(t') $.
			While the Dyson series has recently been used numerically~\cite{kalev_integral-free_2020}, it is challenging to work with because once truncated, it is in general no longer unitary.

			The Magnus series, in contrast, gives $ \mathcal{U}(t, t_0) $ in terms of an exponential of a series of anti-Hermitian operators, specifically
			\begin{align}
				\mathcal{U}(t, t_0) &= \exp\left(\Omega(t, t_0)\right) = \exp\left(\sum_{m = 1}^\infty \Omega_m(t, t_0)\right).
			\end{align}
			As such, the Magnus series explicitly preserves unitarity when truncated~\cite{magnus_exponential_1954}.
			The terms in the Magnus series are integrals over nested time commutators of the anti-Hermitian matrix $ A(t') $, where, in the case of quantum mechanics, $ A(t') = -iH(t') $.
			For instance, the first three terms are~\cite{blanes_magnus_2009}
			\begin{align}
				\Omega_1(t, t_0) &= \int_0^t\text{d}t_1A(t_1),\\
				\Omega_2(t, t_0) &= \frac12\int_{t_0}^t\text{d}t_1\int_{t_0}^{t_1}\text{d}t_2[A(t_1), A(t_2)],\\
				\Omega_3(t, t_0) &= \frac16\int_{t_0}^t\text{d}t_1\int_{t_0}^{t_1}\text{d}t_2\int_{t_0}^{t_2}\text{d}t_3\left([A(t_1), [A(t_2), A(t_3)]] + [A(t_3), [A(t_2), A(t_1)]]\right),
			\end{align}
			where $ [X, Y] = XY - YX $ is the commutator.
			Note that, when truncated to first order, the Magnus expansion $ \mathcal{U}(t, t_0) = \exp\left(\int_{t_0}^t\text{d}t_1\left(-iH(t_1)\right)\right) $ reduces to Equation~\eqref{eq:exp_sol_of_constant}, with $ H(t) $ approximated by its average value over the interval $ [t_0, t] $.


			Convergence of the Magnus series for $ \mathcal{U}(t, t_0) $ is not in general guaranteed, but it is under the condition that $ \int_{t_0}^t\text{d}t_1\|H(t_1)\|_2 < \xi \approx 1.08686870\dots $~\cite{blanes_magnus_2009}.
			Furthermore, each subsequent term in the expansion increases in complexity rapidly.
			For these reasons, the Magnus expansion is generally used as a time-stepping method rather than a single step to solve the complete system.
			Designing a Magnus-based solver is a trade-off between the number of terms used in each expansion, and the number of time-steps used.
			Time stepping is based on the fact that time evolution operators can be split into a product via
			\begin{align}
				\mathcal{U}(t, t_0) &= \mathcal{U}(t, t_{n - 1})\mathcal{U}(t_{n - 1}, t_{n - 2})\cdots\mathcal{U}(t_2, t_1)\mathcal{U}(t_1, t_0).\label{eq:product_of_time_evolution}
			\end{align}
			Each of the time-evolution operators can be approximated by a Magnus expansion.


			Many unitary time-stepping techniques have been developed based on the Magnus expansion~\cite{auer_magnus_2018}.
			Some of these techniques use Gauss-Legendre quadrature sampling and the Baker–Campbell–Hausdorff formula to respectively avoid the integration and time commutators of the Hamiltonian, producing simple expressions that can be used for unitary time-stepping~\cite{blanes_fourth-_2006}.
			
			In particular, we use the commutator free, fourth order method (CF4) from Reference~\cite{auer_magnus_2018}.
			Suppose we wish to evaluate a time-step of $ \mathcal{U}(t + \delta t, t) $ using the CF4 method.
			The Hamiltonians are sampled at times
			\begin{align}
				t_1 &= t + \frac12 \left(1 - \frac{1}{\sqrt{3}}\right)\delta t\quad\text{and}\\
				t_2 &= t + \frac12 \left(1 + \frac{1}{\sqrt{3}}\right)\delta t,
			\end{align} 
			based on the second order Gauss-Legendre quadrature
			\begin{align}
				\overline{H}_1 &= \frac{3 + 2\sqrt{3}}{12}H(t_1) + \frac{3 - 2\sqrt{3}}{12}H(t_2)\quad\text{and}\label{eq:cf4_sample_1}\\
				\overline{H}_2 &= \frac{3 - 2\sqrt{3}}{12}H(t_1) + \frac{3 + 2\sqrt{3}}{12}H(t_2).\label{eq:cf4_sample_2}
			\end{align}
			Then
			\begin{align}
				\mathcal{U}(t + \delta t, t) = \exp(-i\overline{H}_2\,\delta t)\exp(-i\overline{H}_1\,\delta t) + \mathcal{O}(\delta t^5),\label{eq:cf4_implementation}
			\end{align}
			is used to approximate the time-evolution operator.

		\subsubsection{Lie-Trotter based exponentiator}
			Evaluating matrix exponentials is a core part of the integration algorithm.
			Rather than exponentiating the Hamiltonian directly as in Equation~\eqref{eq:cf4_implementation}, \emph{Spinsim} works with the field functions $ \omega_j(t) $.

			For spin-half, the exponentiator is in an analytic form in $ \omega_x, \omega_y $ and $ \omega_z $. 
			For spin-one, an exponentiator based on the Lie-Trotter product formula~\cite{moler_nineteen_2003}
			\begin{align}
				\exp\left( X + Y\right) &= \lim_{n\to\infty} \left(\exp\left(\frac{X}{n}\right) \exp\left(\frac{Y}{n}\right)\right)^n,\label{eq:lie_trotter}
			\end{align}
			is used.
			An advantage of the Lie-Trotter approach is that $ \exp(-iA_k/n) $ has known analytic forms for the Lie algebra basis elements $ A_k $.
			Hence the unitary time-evolution operator is approximated by $ U = \exp(-iH\,\delta t) = T^n + \mathcal{O}(\delta t^3) $, where
			\begin{align}
				T &= \exp\left(\frac{-i \omega_x J_x}{n}\right) \exp\left(\frac{-i \omega_y J_y}{n}\right) \exp\left(\frac{-i \omega_z J_z}{n}\right) \exp\left(\frac{-i \omega_q Q}{n}\right).
			\end{align}
			In fact, commutation relations between the Lie basis operators and the leapfrog splitting method~\cite{barthel_optimized_2020} allow us to instead write 
			\begin{align}
				T &= \exp\left(-i\frac12D_{z,q}\right)\exp(-i\Phi J_\phi)\exp\left(-i\frac12D_{z,q}\right).
			\end{align}
			Here $ z = \omega_z/n $, $q = \omega_q/n $, $ \Phi = \sqrt{\omega_x^2 + \omega_y^2}/n $, $ \phi = \text{arctan}2(\omega_y, \omega_x) $ (where $ \text{arctan}2(y, x) $ is the two argument arctangent~\cite{organick_fortran_1966}), $ D_{z,q} = zJ_z + qQ $ and $ J_\phi = \cos(\phi) J_x + \sin(\phi) J_y $.
			The element-wise analytic form of this is
			\begin{align}
				T &= \begin{pmatrix}
					\left(\cos\left(\frac{\Phi}{2}\right) e^{-iz/2}e^{-iq/6}\right)^2 & \frac{-i}{\sqrt{2}} \sin(\Phi)e^{iq/6}e^{-iz/2}e^{-i\phi} & -\left(\sin\left(\frac{\Phi}{2}\right)e^{iq/6}e^{-i\phi}\right)^2\\
					\frac{-i}{\sqrt{2}} \sin(\Phi)e^{iq/6}e^{-iz/2}e^{i\phi} & \cos(\Phi)e^{i4q} & \frac{-i}{\sqrt{2}} \sin(\Phi)e^{iq/6}e^{iz/2}e^{-i\phi}\\
					-\left(\sin\left(\frac{\Phi}{2}\right)e^{-iq/6}e^{i\phi}\right)^2 & \frac{-i}{\sqrt{2}} \sin(\Phi)e^{iq/6}e^{iz/2}e^{i\phi} & \left(\cos\left(\frac{\Phi}{2}\right) e^{iz/2}e^{-iq/6}\right)^2
				\end{pmatrix}.\label{eq:lie_trotter_4}
			\end{align}

			Matrix exponentiation is completed by raising $ T $ to an integer $ n $.
			To approximate the limit in Equation~\eqref{eq:lie_trotter}, $ n $ must be a large number.
			We choose $ n $ to be of the form $ n = 2^\tau $, as raising matrices to powers of two can be done efficiently by iterative squaring.
			While there are some loose upper bounds for the $ n $ required to reach a desired accuracy for the matrix exponential~\cite{suzuki_generalized_1976}, using these to choose a value of $ n $ can, in practice, cause floating-point errors from over-squaring.
			Instead, we tested the formula on $ 10^5 $ matrices for different values of $ \tau $, and compared them to results given by \texttt{scipy.linalg.expm()}, from the popular python library \emph{SciPy}~\cite{virtanen_scipy_2020}.
			We found that the error minimized after a number of squares of $ \tau = 24 $.
			Thus, by default, $ n = 2^{24} $.

			In addition to those introduced by the Lie-Trotter approximation, we find errors when manipulating matrices close to the identity.
			Because $ n $ is very large, $ T $ is very close to the identity.
			In particular, the limited precision of floating-point numbers is unable to represent the diagonal of $ T $ and its squares accurately.
			Errors appear in two forms.
			One form is that iterative squaring of matrices close to the identity produces floating-point cancellation errors.
			It can be avoided by working with the difference of the matrices from the identity when iteratively squaring.
			That is, if $ A = I + a $, then instead of calculating the square $ S = I + s = A^2 $, we find $ s = (a + 2I)a $.
			The other form of error is that if we use this method, we must initially calculate $ T - I $ accurately, rather than $ T $.
			This cannot be done by simply subtracting the identity from Equation~\eqref{eq:lie_trotter_4}, as doing so would introduce cancellation error.
			It can be avoided by replacing the exponentials and trigonometric functions along the diagonal with specialized implementations of functions such as $ \text{expm}1(x) = \exp(x) - 1 $, designed for this purpose~\cite{hewlett-packard_hp_1994}.
			The identity is added to the result of the iterative residual squaring and the full exponential is returned.
			
			This spin-one exponentiator evolves Hamiltonians spanned by $ J_x $, $ J_y $, $ J_z $ and $ Q $ which is sufficient for three level systems in arbitrary bias fields, but with single-photon coupling.
			An exponentiator capable of evolving an arbitrary spin-one Hamiltonian, for example with different coupling between the lower and upper pairs of states, or with two-photon coupling, is included in the package.
			When assessing accuracy for spin-one problems, we have used the single-photon exponentiator.

			Note that, the methods for both spin-half and spin-one use analytic forms of matrix exponentials so that $ T $ is unitary by construction; $ \mathcal{U}=T^n $ is then also unitary.
			Simulations in \emph{Spinsim} thus maintain unitarity and so conserve probability even over very large numbers of time-steps.

		\subsubsection{Discretization and parallelization}\label{sec:parallelization}
			Frequently we wish to sample the state at times $ t_k $ spaced more coarsely than the integration time-step.
			Consider discrete times $ t_k = t_0 + \Delta t\cdot k $ where $ \Delta t $ is the time-step of the time-series, in contrast to $ \delta t $, the time-step of the integration.
			Denoting $ \psi_k = \psi(t_k) $ and $ \mathcal{U}_k = \mathcal{U}(t_{k}, t_{k-1}) $, the time-series of states $  \psi_k $ and time-evolution operators $ \mathcal{U}_k $ satisfy
			\begin{align}
				\psi_k &= \mathcal{U}_k\psi_{k-1}.\label{eq:integration_compilation}
			\end{align}
			This presents an opportunity for parallelism.
			While $ \psi_k $ depends on $ \psi_{k-1} $, the operators $ \mathcal{U}_k $ depend only on the field functions, and not on previous states $ \psi_{k'} $ or previous operators $ \mathcal{U}_{k'} $.
			Hence, the time-evolution operators $ \mathcal{U}_k $ can be calculated in parallel.

			\emph{Spinsim} splits the full simulation into time intervals $ [t_{k - 1}, t_{k}] $, and calculates time-evolution operators $ \mathcal{U}_k $ for these intervals in parallel on a GPU. 
			When all the $ \mathcal{U}_k $ are calculated, the CPU then multiplies them together (a comparatively less demanding job than calculating them) using Equation~\eqref{eq:integration_compilation} to determine the output samples $ \psi_k $ of the evolving state.

			Beyond discretization for parallelization, we discretize our time-evolution operator into individual integration time-steps.
			Each of the $ \mathcal{U}_k $ is further split into products $ u^k_{L-1} \cdots u^k_0 $ of $ L $ time-evolution operators, now separated by the integration time-step $ \delta t = \Delta t/L $.
			
			To choose an appropriate value of $ \delta t $, users can refer to the accuracy plots in Figures~\ref{fig:benchmark_spin_one_step_error} and~\ref{fig:benchmark_spin_half_step_error} under Section {\ref{sec:accuracy}}. 
			To choose an appropriate value of $ \Delta t $, users must consider three factors.
			For the most benefit out of GPU parallelization, users should choose $ \Delta t $ so that the number of time samples $ K $ is at least as large as the number of GPU cores available for \emph{Spinsim}.
			Users should also make sure that $ \Delta t $ represents a large enough sample rate for the state $ \psi_k $ or expected spin projection $ \langle \overrightarrow{J}\rangle_k $ to be used in their further analysis.
			Finally, if $ K $ is too large, then the GPU will run out of memory on compilation, which will raise an exception. 
			We found that the 10GiB \emph{GeForce RTX 3080} could run at full memory capacity with 60 million sample points for spin-one mode and 120 million sample points for spin-half.


		\subsubsection{Dynamically rotating frame}\label{sec:rotating_frame}
			Hamiltonians which have dominant and slowly-varying terms induce rotations around a primary axis, which is usually chosen to be the quantization axis (which is the $ z $ axis when representing spin operators with the Pauli matrices).
			Transforming from the lab (standard) frame of reference into one rotating around this axis by rotation operator $ R_{\omega_r}(t) = \exp(i\omega_r J_z t) $, the resulting system follows a new Hamiltonian
			\begin{align}
				H_r(t) &= R_{\omega_r}(t) (\omega_x(t) J_x + \omega_y(t) J_y) R_{\omega_r}(-t) + (\omega_z(t) - \omega_r)J_z +\omega_q(t) Q
			\end{align}
			It is a common technique in solving quantum mechanical problems to enter rotating frames, and their more abstract counterparts of interaction pictures~\cite[(p336,338)]{j_j_sakurai_jun_john_modern_2011}.
			This is almost always done to enable the rotating wave approximation (RWA), which is an assumption that the oscillatory components of $ H_r(t) $ on average make minor contributions to time evolution, and can be ignored.
			In some cases, this allows one to obtain analytic, but approximate, solutions to the quantum system.
			Note that the RWA is never invoked by \emph{Spinsim}, as doing this would reduce the accuracy of simulation results; although a problem already expressed in a rotating frame can of course be input by the user if desired.
			In fact, because of the speed increase of \emph{Spinsim} compared to alternatives, users who would have previously simulated their systems using the RWA due to speed concerns may now wish to use \emph{Spinsim} to simulate beyond RWA effects that would previously have been missed.

			Instead we take advantage of rotating frames to increase numerical accuracy.
			By our assumption that $ |\omega_r| \approx |\omega_z(t)| \ge |\omega_x(t)|, |\omega_y(t)|, |\omega_q(t)| $ for the duration we have entered the frame, then the rotating frame Hamiltonian will have a smaller norm than the lab frame Hamiltonian.
			This will yield more accurate results when making discrete time-steps by the integrator, as the distances stepped are smaller.

			A different rotating frame is entered by \emph{Spinsim} for each time interval $ [t_{k - 1}, t_k] $.
			This dynamic application allows for the rotating frame to be used even during sweeps of the dominant term $ \omega_z(t) $ where it is only approximately constant locally.
			Here we choose $ \omega_r = \omega_z(t_{k - 1} + \Delta t/2) $ as an approximate average value of $ \omega_z(t) $ over the duration.
			The calculated time-evolution operator $ \mathcal{U}_k^r $ is then transformed back to the lab frame as
			\begin{align}
				\mathcal{U}_k &= R_{\omega_r}(-\Delta t)\mathcal{U}_k^r.\label{eq:exit_rotating_frame}
			\end{align}
			This functionality can be disabled for systems that do not fit these criteria.

		\subsubsection{Integrator architecture}
			\begin{figure}[h!]
				\centering
				\includegraphics[scale=0.475]{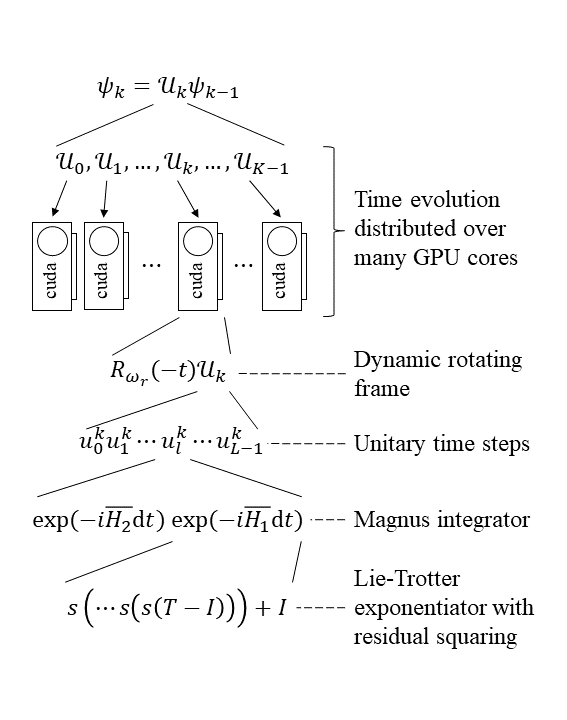}
				\caption{
					The mathematical and computational breakdown of \emph{Spinsim}.
					The procedural problem of evaluating the state time series $ \psi_k $ is broken into the parallel problem of evaluating time-evolution operators $ \mathcal{U}_k $.
					Each of the $ \mathcal{U}_k $ are found independently on a separate GPU thread.
					The problem is brought into a dynamic rotating frame to increase numerical accuracy.
					The $ U_k $ are broken further into time-stepping operators $ u_k $, each of which are evaluated using the Magnus CF4 method.
					Matrix exponentials are evaluated using a Lie-Trotter based method, including a residual squaring technique to avoid over-squaring.
				}
				\label{fig:architecture}
			\end{figure}
			A visual summary of the structure of \emph{Spinsim} is shown in Figure~\ref{fig:architecture}.
			The integrator in the \emph{Spinsim} package calls a kernel (multithreaded function) to be run on a \emph{Cuda} capable \emph{Nvidia} GPU in parallel, with a different thread being allocated to each of the $ \mathcal{U}_k $.
			Alternatively, the kernel can be run in parallel on a multicore CPU if a compatible GPU is not available.
			The kernel returns when each of the $ \mathcal{U}_k $ have been evaluated.

			Each thread begins with initialization of variables, which includes moving into the rotating frame.
			It then starts a loop to find each of the time-stepping operators $ u^k_l $.
			Within the loop, the integrator enters a device function to sample the user-provided field functions $ \omega_j(t) $ (device functions are GPU subroutines, and in practice all device functions here are compiled inline for speed).
			After this, it applies the rotating frame transformation $R_{\omega_r}(t)$, before calculating the Gauss-Legendre weightings $ \overline{H}_1 $ and $ \overline{H}_2 $. Next, the matrix exponentials are taken within another device function.
			Finally, $ u^k_l $ is premultiplied to $ \mathcal{U}^r_k $ (which is initialized to the identity), and the loop continues.
			
			When the loop has finished, $ \mathcal{U}^r_k $ is transformed to $ \mathcal{U}_k $ as in Equation~\eqref{eq:exit_rotating_frame}, and this is returned.
			Once all threads have executed, using Equation~\eqref{eq:integration_compilation}, the state $ \psi_k $ is calculated in a (CPU) \texttt{numba.jit()}ed function from the $ \mathcal{U}_k $ and an initial condition $ \psi_{\text{init}} = \psi(t_0) $.

		\subsubsection{Compilation of integrator}
			The \emph{Spinsim} integrator is constructed and compiled just-in-time (JIT), using the \emph{Numba}~\cite{lam_numba_2015} python package.
			This is achieved via the \texttt{numba.cuda.jit()} decorator, which compiles python functions into \emph{Nvidia Cuda}~\cite{nickolls_scalable_2008} kernels using an LLVM~\cite{lattner_llvm_2004} (Low Level Virtual Machine) compiler.
			The particular device functions used are not predetermined, but are instead chosen based on user input to decide on a closure.
			This technique has multiple advantages.
			First, the field functions $ \omega_j(t) $ are provided by the user as a plain python method.
			Note that this method must be \texttt{numba.cuda.jit()} compatible.
			This allows users to define $ \omega_j(t) $ in a way that compiles and executes fast, does not put many restrictions on the form of the function, and returns the accurate results of analytic functions (compared to the errors seen in interpolation).
			Compiling the simulator also allows the user to set meta-parameters, and choose the features they want to use in a way that does not require experience with the \texttt{numba.cuda} library.
			This was especially useful for running benchmarks comparing integration methods from previous versions of the software to the current one, CF4.
			The default settings should be optimal for most users, although tuning the values of \emph{Cuda} meta-parameters \texttt{max\_registers} and \texttt{threads\_per\_block} could improve performance for different GPUs.
			Third, JIT compilation also allows the user to select a target device other than \emph{Cuda} for compilation, so the simulator can run, using the same algorithm, on a multicore CPU in parallel instead of a GPU.
			In particular, this allows for compatibility with products running \emph{Apple}'s \emph{MacOS 10.14 Mojave} and later, which are incompatible with \emph{Cuda} devices and software.
			
			Customization and execution of the simulator is interfaced through an object of class \texttt{spinsim.Simulator}.
			The \emph{Cuda} kernel is defined as per the user’s instructions on construction of the instance, and it is used by calling the method \texttt{spinsim.Simulator.evaluate()}.
			Users are returned an instance of \texttt{spinsim.Results} including the time, state, time-evolution operator, and expected spin projection.
			Note that, to avoid making unnecessary calculations, the expected spin projection is calculated as a lazy parameter if needed rather than returned by the simulator object.

			The \emph{Spinsim} package is designed so that a single simulator instance can be used to execute many simulations, sweeping through parameters while not needing to be recompiled.
			This is done through the \texttt{sweep\_parameters[]} argument to the user-provided field functions $ \omega_j(t) $.
			First the user must instantiate a \texttt{spinsim.Simulator} object.
			They can then set the value for \texttt{sweep\_parameters[]} for a particular simulation each time \texttt{spinsim.Simulator.evaluate()} is called.
			As \texttt{sweep\_parameters[]} is a \texttt{numpy.ndarray}~\cite{harris_array_2020}, one can use this functionality to sweep many parameters using the same simulator object.
			For a use-case example, an MRI experimentalist might want to simulate many spin systems at different locations $ (x, y) $, in a magnetic field gradient $ \omega_z(t) = x - 2 y $.
			To do this they could choose to set \texttt{sweep\_parameters = [x, y]}, and define \texttt{field\_sample[2] = sweep\_parameters[0] - 2*sweep\_parameters[1]}. 
			This feature is demonstrated with examples in the documentation.

	\subsection{Quality control}
		\subsubsection{Evaluation of accuracy}\label{sec:accuracy}
			All accuracy benchmarks were run in \emph{Cuda} mode, on the desktop computer with the \emph{Ryzen 7 5800X} and \emph{GeForce RTX 3080}, which from Figure~\ref{fig:benchmark_device_aggregate} are the fastest CPU and GPU from the devices tested.
			Note that these benchmarks do not take into account the amount of time required to JIT compile simulation code for the first time (order of seconds), as we want to look at this in the limit of running many simulations while sweeping through a parameter that differs in each one.
			
			Benchmarks were performed using \texttt{neural-sense.sim.benchmark} (where \texttt{neural-sense}~\cite{tritt_neural_2020} is the quantum sensing package that \emph{Spinsim} was written for).
			This simulation involves continuously driving transitions in the system for $ 100\,\text{ms} $, while exposing it to a $ 1\,\text{ms} $ pulsed signal that the system should be able to sense.
			The system has the Hamiltonian
			\begin{align}
				H(t) = \omega J_z + 2\Omega\cos(\omega t)J_x + \Omega_p \text{sinp}(\Omega(t - t_p)) J_z,\label{eq:neural_pulse}
			\end{align}
			where $ \omega = 2\pi\cdot700\,\text{kHz} $, $ \Omega = 2\pi\cdot1\,\text{kHz} $,$ \Omega_p = 2\pi\cdot70\,\text{Hz} $, $ t_p = 233\,\text{ms} $ and $ \text{sinp} $ is a single cycle of a sine wave.

			We first wanted to test the accuracy of the different integration techniques for various integration time-steps.
			Here we wanted to test the advantages of using a Magnus-based integration method.
			Accuracy was calculated by taking the quantum state simulation evaluations of a typical quantum sensing experiment and finding the root mean squared (RMS) error from a baseline simulation run by \texttt{scipy.integrate.ivp\_solve()} as part of the \emph{SciPy} python package, via
			\begin{align}
				\epsilon &= \frac{1}{K}\sqrt{\sum_{k = 0}^{K - 1}\sum_{m_j = -j}^j|\psi_{k, (m_j)} - \psi_{k, (m_j)}^{\textrm{baseline}}|^2},\label{eq:error}
			\end{align}
			where $ j \in \{\frac12, 1\} $ is the spin quantum number of the system.
			This baseline was computed in 2.4 hours (in comparison to order of $ 100\,\text{ms} $ these \emph{Spinsim} tests were executed in), and was also used for comparisons to other software packages.

\begin{figure}[h!]
				\begin{subfigure}[b]{0.475\textwidth}
					\includegraphics[scale=0.475]{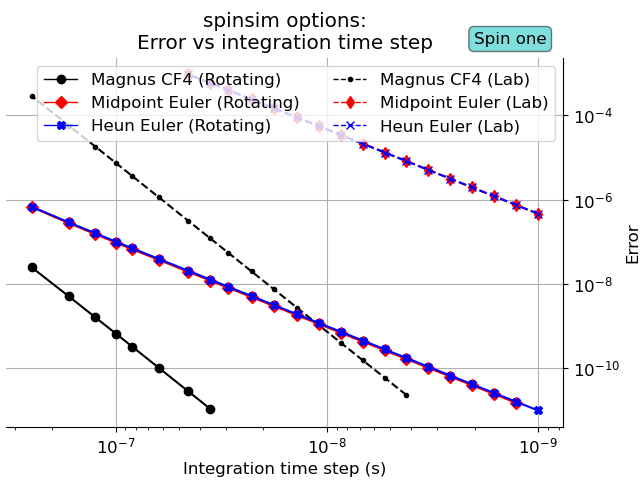}
					\caption{}
					\label{fig:benchmark_spin_one_step_error}
				\end{subfigure}
				\hfill
				\begin{subfigure}[b]{0.475\textwidth}
					\includegraphics[scale=0.475]{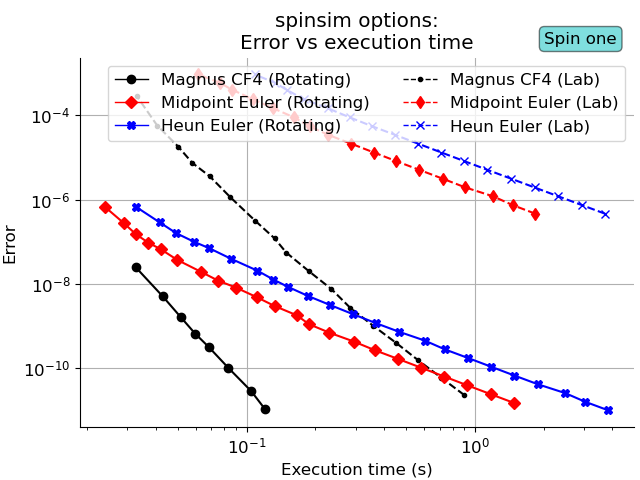}
					\caption{}
					\label{fig:benchmark_spin_one_execution_error}
				\end{subfigure}
				\vfill
				\begin{subfigure}[b]{0.475\textwidth}
					\includegraphics[scale=0.475]{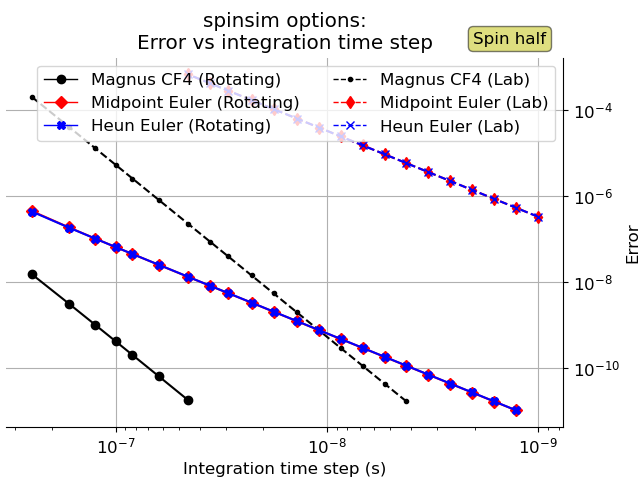}
					\caption{}
					\label{fig:benchmark_spin_half_step_error}
				\end{subfigure}
				\hfill
				\begin{subfigure}[b]{0.475\textwidth}
					\includegraphics[scale=0.475]{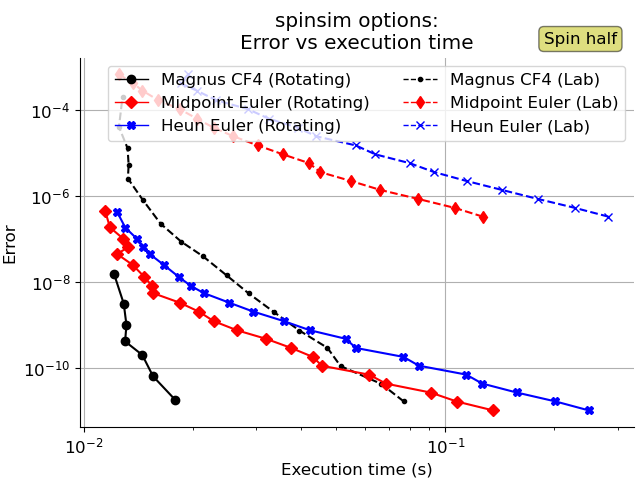}
					\caption{}
					\label{fig:benchmark_spin_half_execution_error}
				\end{subfigure}
				\caption{Speed and accuracy of the spin-one and spin-half options of \emph{Spinsim}.
				A simulation of a typical sensing experiment for our lab was run for every integration time-step, for each of the possible integration techniques (Magnus-based commutator free 4, and two Euler methods).
				In the simulation, transitions are continuously driven in the spin system for a duration of $ 100 \,\text{ms} $, and an additional small signal is injected for $ 1\,\text{ms} $.
				See Equation~\eqref{eq:neural_pulse}.
				Each technique was tested while both using and not using a transformation into a rotating frame.
				Both execution time and error were recorded for each of the simulations.
				Error is RMS error compared to a long running \emph{SciPy} baseline.}
				\label{fig:benchmark_spin}
			\end{figure}

			Error vs time-step plots are shown in Figures~\ref{fig:benchmark_spin_one_step_error} and~\ref{fig:benchmark_spin_half_step_error}.
			In order to find how long a simulation method takes to complete for a given accuracy, we also measured the execution time for each of the simulations.
			These plots are shown in Figures~\ref{fig:benchmark_spin_one_execution_error} and~\ref{fig:benchmark_spin_half_execution_error}.
			The former sets of plots will be of interest to users choosing an appropriate time-step for simulations, and the latter for estimating the execution time of such simulations.
			In all of these comparisons, errors above $ 10^{-3} $ were counted as a failed simulation, as quantum states cannot be arbitrarily far away from each other given that they are points on a unit complex sphere.
			Errors below $ 10^{-11} $ were also excluded, as this was the order of magnitude of the errors in the reference simulation.

			The integration techniques tested were the Magnus-based CF4, as well as two Euler-based sampling methods.
			A midpoint Euler method was chosen as the simplest (and fastest for a given integration time-step) possible sampling method, whereas a Heun-Euler sampling method was used as a comparison to previous versions of this code.
			These are respectively labelled as the modified Euler method and the improved Euler method in Reference~\cite[p328]{suli_introduction_2003}.
			We benchmarked these methods both while using and not using the rotating frame option.
			This was done separately for spin-one and spin-half systems, to ensure they both yield accurate results.

			From Figure~\ref{fig:benchmark_spin}, we find that overall, the results that \emph{Spinsim} gives are accurate to those of \emph{SciPy}.
			Figures~\ref{fig:benchmark_spin_one_step_error}, and~\ref{fig:benchmark_spin_half_step_error} show that using the Magnus-based integration method is up to 3 orders of magnitude more accurate when compared the Euler-based methods.
			Also, using the rotating frame increased the accuracy here by 4 orders of magnitude for any individual integration method.
			When comparing speed vs accuracy in Figures~\ref{fig:benchmark_spin_one_execution_error}, and~\ref{fig:benchmark_spin_half_execution_error}, the advantage that CF4 gives in terms of accuracy far outweighs its slower execution speed when compared to midpoint Euler methods.

		\subsubsection{Comparison to alternatives}
			We ran the same error and execution time benchmarks on some alternative packages, listed in Table~\ref{tab:external}, to compare \emph{Spinsim}'s performance to theirs.
			\begin{table}[h!]
				\caption{The software packages used for and verification of \emph{Spinsim}.}
				\label{tab:external}
				\begin{tabular}{l|l|p{6.5cm}}
					\textbf{Software package}								&\textbf{Function}					&\textbf{Details}\\
					\hline
					Spinsim													&\texttt{Simulator()}				&The best performing \emph{Spinsim} configuration, using the CF4 integrator and the rotating frame mode. This was run both on CPU and GPU.
					\\
					\hline
					QuTip~\cite{johansson_qutip_2013}						&\texttt{sesolve()}					&The Schr\"odinger equation solver from the popular \emph{python} quantum mechanics library, \emph{QuTip}.
					This was chosen as a comparison to a specially designed solver used within the physics community for this application.
					Some quantum mechanics simulation packages such as the recent \emph{scqubits}~\cite{groszkowski_scqubits_2021} use \emph{QuTip} for calculating time evolution.
					Like \emph{Spinsim}, \emph{QuTip} allows users to sample from compiled functions, and uses parallelization.\\
					\hline
					Mathematica~\cite{wolfram_research_inc_mathematica_2020}	&\texttt{NDSolve()}					&A generic ODE solver from the \emph{Mathematica} software.
					This was chosen as it has been historically popular with our lab group for simulating quantum sensing experiments.\\
					\hline
					SciPy~\cite{virtanen_scipy_2020}							&\texttt{integrate.ivp\_solve()}	&A generic ODE solver from the popular \emph{python} scientific computing library.
					This was chosen as a comparison to a generic solver from within the python ecosystem.
				\end{tabular}
			\end{table}
			To obtain simulation results of different accuracies, the step sizes of the alternative integrators were limited to a maximum value.
			In some cases, the maximum number of steps was modified in some cases to allow for the smaller step sizes.
			Apart from that, the integrator settings were left untouched from the default values, as a representation of what a user would experience using a generic solver for spin system problems.

			Similarly to the internal \emph{Spinsim} benchmarks (see Section {\ref{sec:accuracy}}), the expected spin projection was evaluated in each case, but only the states were compared to calculate a relative error via Equation~\eqref{eq:error}.
			Again, we used the longest running \emph{SciPy} simulation as a baseline for comparison, as the accuracy of \emph{Mathematica} plateaus at small time-steps.
			Functions used for sampling were compiled to \emph{LLVM} and \emph{cython} in \emph{Spinsim} and \emph{QuTip} respectively.
			In both cases, the time taken to complete a simulation was measured using a second simulation using the already compiled functions, to represent the use-case of sweeping through many simulations.

			The speed of only one simulation was measured for each benchmark.
			However, it might be possible to increase the average speed of many benchmarks from \emph{Mathematica} and \emph{SciPy} packages by using multithreading to run multiple simulations at a time.
			We attempted doing 8 way multithreading (on an 8 core \emph{Ryzen 7 5800X}) with \emph{Mathematica}, but the solver crashed due to insufficient RAM (of $ 32\,\text{GiB} $).
			Multithreading was not attempted using \emph{SciPy}, due to the fact that running the full set of benchmarks of only a single simulation per integration time-step already consumes over 11 hours of computational time.
			Regardless, for a fair comparison, both \emph{Mathematica} and \emph{SciPy} results are plotted with an artificial reduction in execution time by a factor of 8 (dotted line in Figures~\ref{fig:benchmark_spin_one_execution_error} and~\ref{fig:benchmark_spin_half_execution_error}), which is an upper bound for the speed increase that could be obtained by running them parallel on an 8 core processor.
			Simulations from \emph{Spinsim} and \emph{QuTip} automatically run multithreaded, so this comparison is not plotted for these packages.

\begin{figure}[h!]
				\centering
				\includegraphics[scale=0.475]{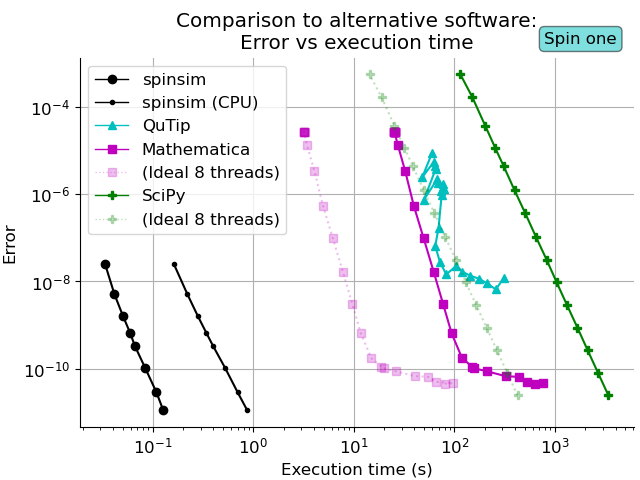} 
				\caption{Speed vs accuracy of two alternative integration packages.
				A simulation of a typical sensing experiment for our lab was run for every integration time-step, for each of alternative packages (\texttt{qutip.sesolve()} from \emph{QuTip}, \texttt{NDSolve()} from \emph{Mathematica}, and \texttt{scipy.integrate.ivp\_solve()} from \emph{SciPy}).
				In the simulation, transitions are continuously driven in the spin system for a duration of $ 100\,\text{ms} $, and an additional small signal is injected for $ 1\,\text{ms} $.
				See Equation~\eqref{eq:neural_pulse}.
				Both execution time and error were recorded for each of the simulations.
				Error is RMS error compared to a long running \emph{SciPy} baseline.
				The \emph{Mathematica} and \emph{SciPy} results are also shown with a speed up by a factor of 8 to represent the upper bound of hypothetical parallelization across an 8 core CPU.}
				\label{fig:benchmark_external}
			\end{figure}

			From Figure~\ref{fig:benchmark_external}, for any given error tolerance, \emph{Spinsim} is over 3 orders of magnitude faster than \emph{Mathematica} and \emph{QuTip}, and 4 orders of magnitude more accurate than \emph{SciPy}.
			In practice, this means that a 25 minute \emph{SciPy} simulation is reduced to 50ms, and a three week long \emph{SciPy} batch simulation of 1000 separate systems (a realistic situation for testing quantum sensing protocols) would take less than one minute in \emph{Spinsim}.
			Note that while the three week long \emph{SciPy} example could be parallelized over a computer cluster, saving some, \emph{Spinsim} allows users to avoid the need for using clusters and, given that a single \emph{SciPy} simulation still takes 25 minutes to run, the \emph{Spinsim} batch would still be faster.

		\subsubsection{Parallelization performance}
\begin{figure}[htbp!]
				\centering
				\includegraphics[scale=0.6]{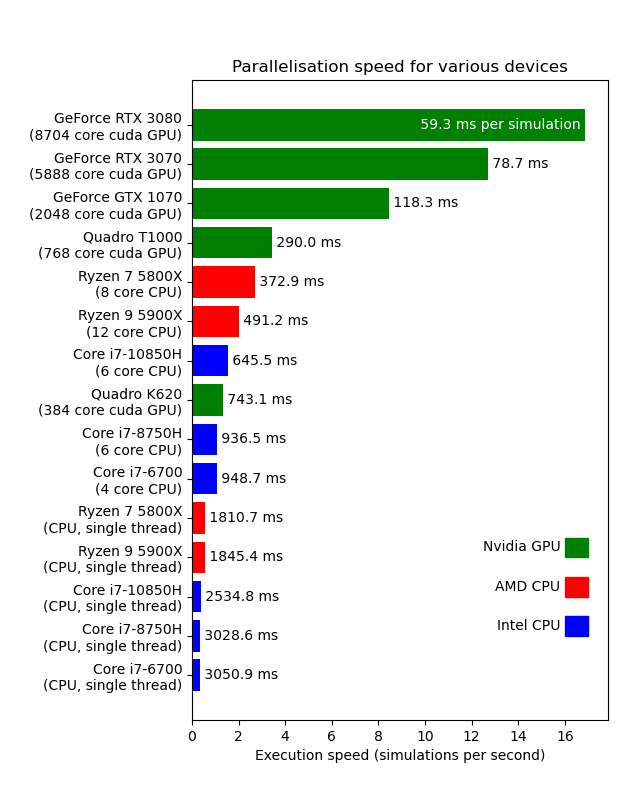}
				\caption{Evaluation speed of a simulation of a typical spin-one sensing experiment for our lab on both CPUs and GPUs.
				Integration time-step is set to $ 100\,\text{ns} $.
				Transitions are continuously driven in the spin system for a duration of $ 100\,\text{ms} $, and an additional small signal is injected for $ 1\,\text{ms} $.
				See Equation~\eqref{eq:neural_pulse}.
				Evaluation time is determined by an average of 100 similar simulations for each device, where each individual simulation varies in dressing amplitude (transition frequency).}
				\label{fig:benchmark_device_aggregate}
			\end{figure}

			Once the algorithm behind \emph{Spinsim} was developed, we wanted to check its execution speed while running on various devices.
			The main reason for this test was to quantify the speed increase of parallelization by comparing execution speeds on highly parallel devices (being GPUs), and highly procedural devices (being CPUs).
			Speed benchmarks were performed using \texttt{neural-sense.sim.benchmark}, by comparing the evaluation speed of typical spin-one sensing experiments on different devices.
			Results are of these benchmarks are shown in Figure~\ref{fig:benchmark_device_aggregate}.
			The integration code was compiled by \texttt{numba} for multicore CPUs, CPUs running single threaded, and \emph{Nvidia Cuda} compatible GPUs.
			The compiled code was then run on different models of each of these devices.
			These test devices are given in Table~\ref{tab:devices}.

			\begin{table}[h!]
				\caption{Devices used in the parallelization speed test.
				These devices are part of individual computers, which are separated here by horizontal lines.}
				\label{tab:devices}
				\begin{tabular}{l|l|l|l|l}
					\textbf{Name}	&\textbf{Device}&\textbf{RAM (GiB)}	&\textbf{Cores}	&\textbf{Cooling}\\
					\hline
					Core i7-6700	&Intel CPU		&16					&4				&Air\\
					Quadro K620		&Nvidia GPU		&2					&384			&Air\\
					\hline
					Core i7-8750H	&Intel CPU		&16					&6				&Air\\
					GeForce GTX 1070&Nvidia GPU		&8					&2048			&Air\\
					\hline
					Core i7-10850H	&Intel CPU		&32					&6				&Air\\
					Quadro T1000	&Nvidia GPU		&4					&768			&Air\\
					\hline
					Ryzen 9 5900X	&AMD CPU		&32					&12				&Air\\
					GeForce RTX 3070&Nvidia GPU		&8					&5888			&Air\\
					\hline
					Ryzen 7 5800X	&AMD CPU		&32					&8				&Liquid\\
					GeForce RTX 3080&Nvidia GPU		&10					&8704			&Air\\
				\end{tabular}
			\end{table}

			The results in Figure~\ref{fig:benchmark_device_aggregate} show the benefit to using parallelization when solving a spin system problem.
			Moving from the 6 core \emph{Core i7-8750H} CPU to the 12 core \emph{Ryzen 9 5900X} CPU doubles the execution speed, as does moving from the 384 core \emph{Quadro K620} GPU to the 768 core \emph{Quadro T1000} GPU.
			So, in these cases, performance scales in proportion to thread count.
			Moving from a single core processor to a high end GPU increases performance by well over an order of magnitude on three of the five computers used for testing.
			Even the low end \emph{Quadro K620} was an improvement over the \emph{Core i7-6700} used by the same computer.
			Execution speed vs number of cuda cores starts to plateau as the number of cores increases.
			This happens because the time it takes to transfer memory from RAM to VRAM (dedicated graphics memory), which is independent on the number of cores of the GPU, becomes comparable to the execution time of the simulator logic.
			However, there is still a large improvement from using the high end \emph{GeForce RTX 3070} to the \emph{GeForce RTX 3080}, with the latter simulating the experiment almost twice as fast as it would take to run the simulated experiment in the real world.

			Surprisingly, we found that the \emph{Ryzen 7 5800X} 8 core CPU was able to execute the benchmark faster than the \emph{Ryzen 9 5900X} 12 core CPU.
			This can be explained by the fact that the \emph{Ryzen 7 5800X} was cooled by liquid rather than air, meaning it was likely able to boost to a higher core clock, and resist thermal throttling.

			Users can view Figure~\ref{fig:benchmark_device_aggregate} to decide on how much their current or future hardware will be able to take advantage of \emph{Spinsim}'s parallelism.
			Another factor not shown in the plot is that, in practice, running highly code parallel code on a CPU on a personal computer will severely limit the responsiveness of other applications, as it can utilize the entire CPU (as it should).
			In contrast, this does not happen when running a GPU based program, as it requires very little CPU utilization to function.
			This can be convenient when running simulations on a personal laptop or desktop, as other work on the computer does not have to halt while simulations are being run.

		\subsubsection{Testing}
			During the accuracy tests, it was confirmed that all possible modes of \emph{Spinsim} agree with a baseline \emph{SciPy} simulation, as close as the user wants up to the error of $ 10^{-11} $ of that baseline.
			Therefore, \emph{Spinsim} can be trusted for accuracy.
			The Lie Trotter matrix exponentiator was tested separately from the full system, as well as benchmarked separately against \texttt{scipy.linalg.expm()} from \emph{SciPy}.
			These tests and benchmarks were run as part of the \texttt{neural\_sense} package.
			The simulator has also been adopted by members of our lab, who have given advice on user experience.

			The kernel execution was profiled thoroughly, and changes were made to optimize VRAM and register usage and transfer.
			This was done specifically for the development hardware of the \emph{GeForce GTX 1070}, so one may get some performance increases by changing some GPU specific meta parameters when instantiating the \texttt{spinsim.Simulator} object.

			A good way to confirm that \emph{Spinsim} is functioning properly after an installation is to run the tutorial code provided and compare the outputs.
			Otherwise, one can reproduce the benchmarks shown here using \texttt{neural\_sense.sim.benchmark}.

\section{Availability}
	\subsection*{Operating system}
	Developed and tested on \emph{Windows 10}.
	CPU functionality tested on \emph{MacOS 10.16 Big Sur} (note that \emph{MacOS 10.14 Mojave} and higher is not compatible with \emph{Cuda} hardware and software).
	The package (including \emph{Cuda} functionality) is in principle compatible with \emph{Linux}, but functionality has not been tested.

	\subsection*{Programming language}
	Python (3.7 or greater)

	\subsection*{Additional system requirements}
	To use the (default) \emph{Nvidia Cuda} GPU parallelization, one needs to have a \emph{Cuda} compatible \emph{Nvidia} GPU~\cite{nvidia_cuda_2012}.
	For \emph{Cuda} mode to function, one also needs to install the \emph{Nvidia Cuda} toolkit~\cite{nvidia_cuda_2013}.
	If \emph{Cuda} is not available on the system, the simulator will automatically parallelize over multicore CPUs instead.

	\subsection*{Dependencies}
	numba (0.50.1 or greater)\\
	numpy (1.19.3)\\
	matplotlib (for example code, 3.2)\\
	neuralsense (for benchmark code)

	\subsection*{Software location:}

	\subsubsection*{Archive}

	Name: Monash Bridges\\
	Persistent identifier: 10.26180/13285460\\
	Licence: Apache 2.0\\
	Publisher: Alex Tritt\\
	Version published: 1.0.0\\
	Date published: 2022-04-11\\

	\subsubsection*{Code repository}

	Name: GitHub\\
	Persistent identifier: https://github.com/alexander-tritt-monash/spinsim\\
	Licence: BSD 3 Clause\\
	Date published: 2020-11-18\\

	\subsection*{Languages}

	English.
\section{Reuse potential}

	\subsection{Applications}\label{sec:applications}
		The \emph{Spinsim} package will be useful for any research group needing quick, accurate, and/or large numbers of simulations involving spin-half or spin-one systems.
		In particular, the package can simulate a system with complicated (i.e.not just hard pulses), analytic or time series driving fields, with no approximations like the RWA required, at a high time resolution. 
		This includes simulating a Hamiltonian of an arbitrary number of tones, of any polarization, that can be modulated by frequency, amplitude and/or phase.
		We stress that many application specific simulators are designed for the paradigm of using the RWA and/or hard pulse approximations~\cite{suzuki_qulacs_2021, teske_qopt_2021, isakov_simulations_2021}, which both sacrifice accuracy for the sake of speed.
		While this is viable for some areas of physics research, when designing quantum sensing protocols we need to take all effects of the spin system and surrounding lab environment into account.
		Thus, \emph{Spinsim}'s primary use is to fill in this gap of simulators for use of developing new quantum sensing protocols with spin-half and spin-one systems, and so should be useful to many within the field of quantum sensing.
		Moreover, \emph{Spinsim} is fast and accurate enough that there is no need to simplify the problem to compromise accuracy for speed.
		
		The package was written for the context of testing a particular ultracold atom magnetic sensing protocol design.
		This project aims to be able to measure neural signals using ultracold atoms.
		Electrical pulses made by neurons are currently measured using electrical probes~\cite{mitterdorfer_potassium_2002}, which is intrusive and damages the cells.
		We, and others in the field~\cite{barry_optical_2016, xu_wavelet-based_2016, parashar_axon_2020, webb_high_2021} instead propose to sense the small magnetic fields that these electrical currents produce.
		Rubidium ultracold atom clouds can potentially be made sensitive enough to these small magnetic waveforms that we can use them as sensors.
		The \emph{Spinsim} package was written to simulate possible measurement protocols for this, showing the behavior of the array of spin-one atoms interacting with the magnetic fields of the neurons, control signals, and the lab environment.


		Aside from cold atom based sensors, \emph{Spinsim} could be used to simulate sensors made from nitrogen vacancy centres (NVCs).
		These are spin-one structures found in diamond doped with Nitrogen atoms.
		Similar to ultracold atoms, NVCs can be placed and addressed in 2D arrays in order to take many samples in one measurement.
		As work first started on \emph{Spinsim}, Parashar et al~\cite{parashar_axon_2020} published results on simulation experiments of magnetic neural pulse sensing using NVCs.
		This is a field \emph{Spinsim} could be useful in.


		With some restrictions of not being able to model mixed states, \emph{Spinsim} could be used for simulations in various areas of NMR.
		There are many atomic nuclei with spins of half (eg protons, Carbon 13) and, and some that have spins of one (eg Lithium 6, Nitrogen 14)~\cite{fuller_nuclear_1976}, which, if relaxation and interactions between systems are not important for the application, \emph{Spinsim} could be used to simulate for spectroscopy experiments, for example.
		The inclusion of a quadrupole operator means that \emph{Spinsim} should be able to simulate nuclear quadrupole resonance (NQR) spectroscopy for spin-one nuclei~\cite{bain_nqr_2004}, such as Nitrogen 14, provided a suitable coordinate system is chosen.
		This technique measures energy level differences between levels split by electric field gradients, rather than static magnetic bias fields.
		Another possible use-case could be for magnetic resonance imaging (MRI) simulation and pulse sequence design.
		MRI uses measures the response of spins of an array of spin-half protons to a spatially varying pulse sequence~\cite{mckinnon_physics_1998}, which essentially just corresponds to many separate \emph{Spinsim} simulations of spins at different positions in space.
		This package offers some advantages over state of the art simulators in the field~\cite{kose_fast_2019}, with its use of quantum mechanics over classical mechanics, and its absence of rotating wave approximations, its parametrized pulse sequence definitions and geometric integrator.

		The field of Quantum Optimal Control (QOC) requires many simulations in order to optimize a pulse sequence to be robust to the noisy environment of the system, using machine learning techniques~\cite{dalgaard_fast_2021}.
		The ability to automatically differentiate~\cite{griewank_who_2012} the simulator is important for this application.
		This is not currently a feature of \emph{Spinsim}.
		However, it could be added in the future, especially since it might also be helpful for some quantum sensing applications.

	\subsection{Support}
		Documentation for \emph{Spinsim} is available on \href{https://spinsim.readthedocs.io/en/latest/}{\emph{Read the Docs} (https://spinsim.readthedocs.io/en/latest/)}.
		This documentation contains a thorough tutorial of examples on how to use the package, and installation instructions.
		
		For direct support with the \emph{Spinsim} package, one can open an issue in the \emph{github} repository.
		One can also use this contact to suggest extensions to the package.
		\emph{Spinsim} is planned to be used and maintained by the Monash University School of Physics \& Astronomy spinor BEC lab into the future.

\section{Acknowledgements}

Thank you to past and present members the Monash University School of Physics \& Astronomy spinor BEC lab group. In particular, Chris Bounds, Hamish Taylor, Travis Hartley, and Sam White, who have been using \emph{Spinsim} to simulate new ideas for quantum sensing protocols, and have given useful feedback of their user experience with the package.

\section{Funding statement}

AT acknowledges support through an Australian Government Research Training Program Scholarship.\\
JS is the recipient of an Australian Research Council Discovery Early Career Researcher Award (project number DE210101056) funded by the Australian Government.\\
LDT acknowledges funding from the Australian Research Council Linkage Project (project number LP200100082).

\section{Competing interests}

The authors declare that they have no competing interests.

\bibliography{spinsim}

\end{document}